\documentclass[conference]{IEEEtran}
\IEEEoverridecommandlockouts
\usepackage{cite}
\usepackage{amsmath,amssymb,amsfonts}
\usepackage{algorithmic}
\usepackage{graphicx}
\usepackage{textcomp}
\usepackage{xcolor}
\def\BibTeX{{\rm B\kern-.05em{\sc i\kern-.025em b}\kern-.08em
    T\kern-.1667em\lower.7ex\hbox{E}\kern-.125emX}}
    
\DeclareRobustCommand*{\IEEEauthorrefmark}[1]{\raisebox{0pt}[0pt][0pt]{\textsuperscript{\footnotesize #1}}}

\begin{document}

\title{Carrier Leakage Suppression and Power Difference Equalization for TFDMA Coherent PON
\thanks{This work was supported in part by National Key R\&D Program of China under Grant 2023YFB2905700, in part by National Natural Science Foundation of China under Grants 62371207 and 62005102, in part by Young Elite Scientists Sponsorship Program by CAST under Grant 2023QNRC001, in part by the Guangzhou Basic and Applied Basic Research Foundation under Grant 2025A04J5117, and in part by Hong Kong Research Grants Council GRF under Grant 15231923.}}

\author{\IEEEauthorblockN{Ji Zhou\IEEEauthorrefmark{1,*}, Haide Wang\IEEEauthorrefmark{2}, Liangchuan Li\IEEEauthorrefmark{1,\dag}, Changyuan Yu\IEEEauthorrefmark{3}, and Xiangjun Xin\IEEEauthorrefmark{1}}
\vspace{6pt}
\IEEEauthorblockA{\IEEEauthorrefmark{1} Aerospace and Informatics Domain, Beijing Institute of Technology, Zhuhai, 519088, China}
\IEEEauthorblockA{\IEEEauthorrefmark{2} School of Cyber Security, Guangdong Polytechnic Normal University, Guangzhou, 510665, China}
\IEEEauthorblockA{\IEEEauthorrefmark{3} Department of Electrical and Electronic Engineering, The Hong Kong Polytechnic University, Hong Kong, China}
\vspace{6pt}
\IEEEauthorblockN{\IEEEauthorrefmark{*} zhouji@bitzh.edu.cn and \IEEEauthorrefmark{\dag} liliangchuan@bitzh.edu.cn}
}
\maketitle

\begin{abstract}
In this invited talk, we demonstrate the carrier leakage suppression and power difference equalization based on semiconductor optical amplifier (SOA) for TFDMA-based coherent PON, and propose a nonlinearity compensation algorithm to deal with SOA-induced nonlinear distortion.
\end{abstract}


\section{Introduction}
\begin{figure}[!t]
\centering
\includegraphics[width = \linewidth]{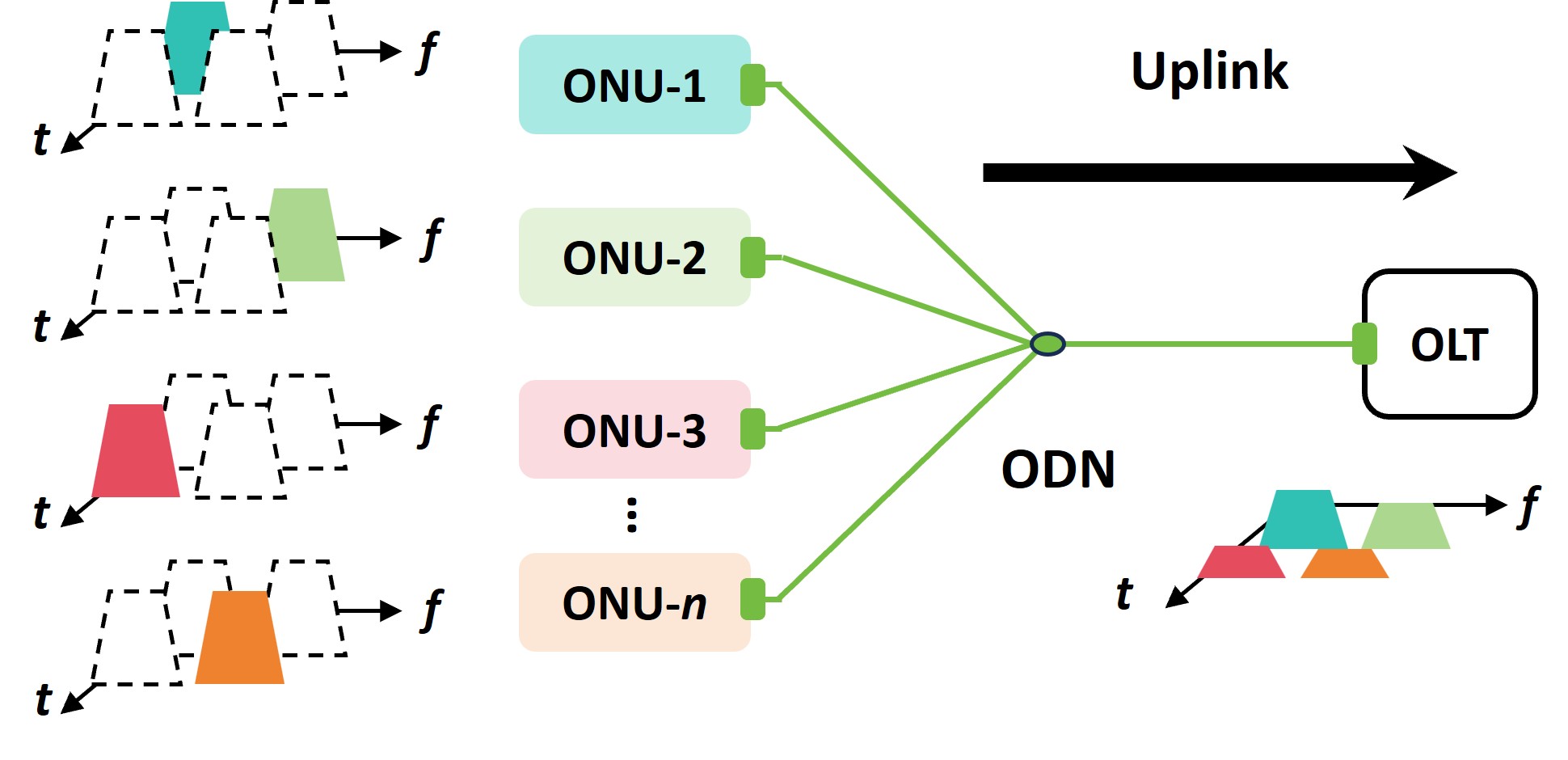}
\caption{The uplink of the coherent PON based on TFDMA when the link losses between the ONUs and the OLT are unequal.}
\label{Fig_1}
\end{figure}

Following the evolutionary trajectory of International Telecommunication Union Telecommunication Standardization Sector (ITU-T) standards, Beyond 50Gb/s passive optical network (B50G-PON) is projected to achieve an access rate of 200Gb/s \cite{jia2025coherent, xing2024low, zhang2022coherent}. Conventional intensity modulation/direct detection (IM/DD) architectures face fundamental challenges in simultaneously satisfying both the access rate and optical power budget requirements. Coherent optical technology has emerged as a viable candidate for B50G-PON, enabling the concurrent achievement of 200Gb/s access rates and a 35dB optical power budget \cite{wei2026dynamic, guo2025flexible, yan2025next}. Meanwhile, coherent PON enables high-capacity, low-latency, and massive-connectivity time-frequency division multiple access (TFDMA) \cite{zhang2020rate, zhou2024flexible, xing2023first}. Many previous studies have addressed two critical issues: burst-mode digital signal processing (BM-DSP) in coherent optical TDMA (CO-TDMA)\cite{zhang2020efficient, zhou2025burst, wang2023fast} and coherent transceiver impairment compensation in coherent optical FDMA (CO-FDMA)\cite{zhou2024iq, wang2024training, zhao2025preamble}. However, coherent PON still faces numerous issues that urgently need addressing. Fig. \ref{Fig_1} shows the uplink of the coherent PON based on TFDMA when the link losses between the optical network units (ONUs) and the optical line terminal (OLT) are unequal. The received signals from different ONUs exhibit different powers, and processing TFDMA signals with such power disparities poses a significant challenge.

For the previous TDMA-based IM/DD PON, ONU can fast turn the optical signal on at its time slots and turn it off at the other time slots by turning on/off directly modulated laser (DML), respectively. Meanwhile, the burst-mode trans-impedance amplifier (BM-TIA) at the OLT can fast adjust the gains to amplify the received burst signals with different link losses\cite{zhang2020progress, nesset2025progress, zhou2025real}. However, in the coherent PON, the laser of the coherent optical transmitter at the ONU cannot be fast turned off and on. When the modulated signal is turned off at the other time slots, the output power of the coherent optical transmitter is not zero due to the limited extinction ratio, which influences the performance of the other ONUs. The high-extinction-ratio optical modulator should be used to suppress the carrier leakage at the time slots of the other ONUs\cite{deng2026silicon}. Semiconductor optical amplifier (SOA) at the ONU was proposed to suppress leakage and equalize power for TDMA \cite{vijayan2025480}. The $N$ high-extinction-ratio optical modulators at the ONUs and one BM-TIA at the OLT may be more cost-effective than the $N$ SOAs at the ONUs where $N$ is the number of ONUs. However, the BM-TIA cannot amplify the subcarrier signals of FDMA with power difference for the TFDMA. SOA at the ONU can simultaneously suppress carrier leakage for TDMA and equalize power difference for FDMA, which seems to be the only option available for TFDMA. 

This invited talk is based on our previous works \cite{zhou2026higher, Chen2026SOA}. We experimentally demonstrate the carrier leakage suppression and power difference equalization based on SOA for TFDMA-based coherent PON, and propose a burst-mode nonlinearity compensation (BM-NLC) algorithm to deal with SOA-induced nonlinear distortion. 

\section{Experimental Setups and Results}
\begin{figure}[!t]
\centering
\includegraphics[width = \linewidth]{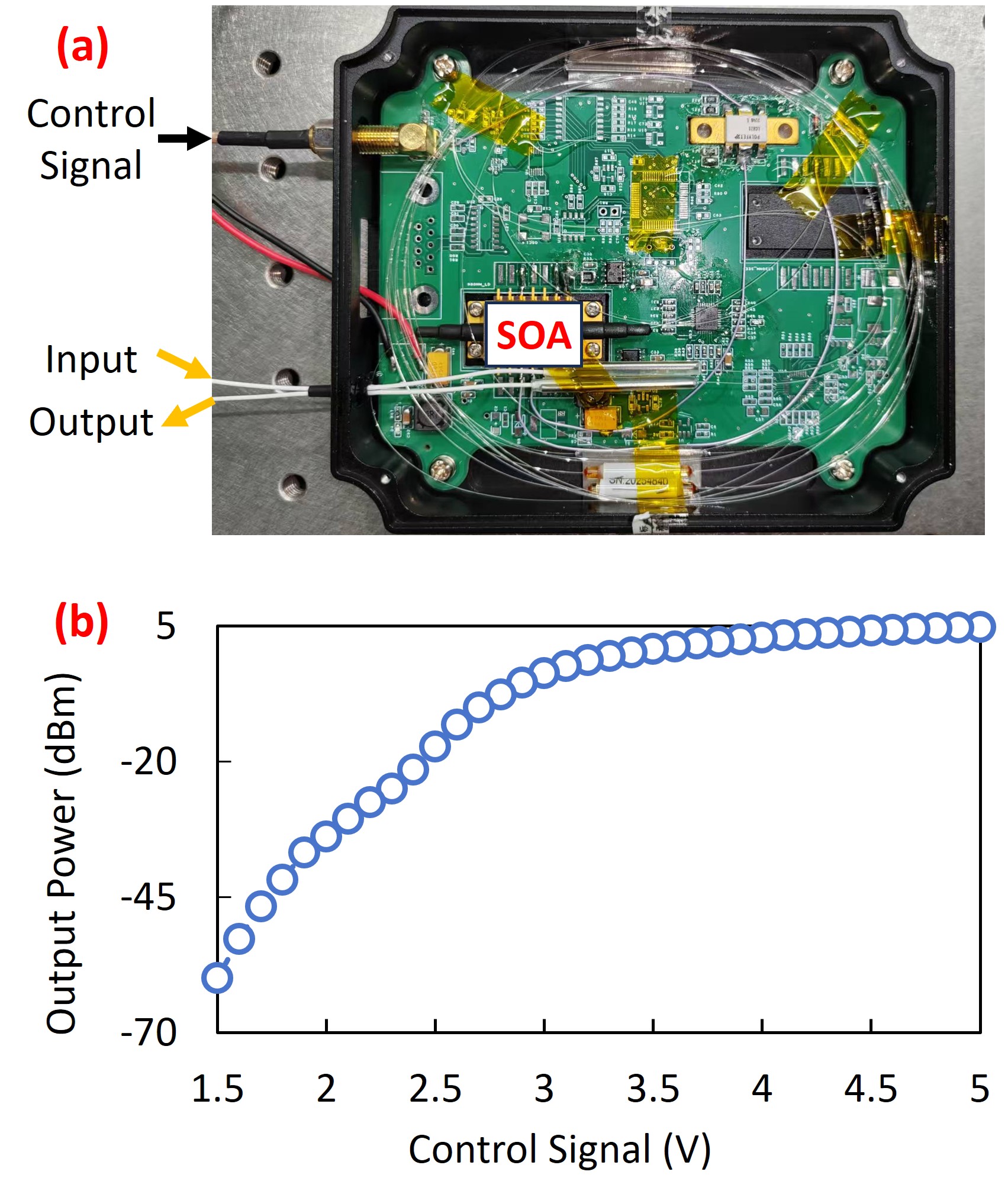}
\caption{(a) SOA evaluation board with adjustable gain based on the control signal; (b) The output power versus the voltage of the control signal for the SOA with an input power of $-$13.5 dBm.}
\label{Fig_2}
\end{figure}

Figure \ref{Fig_2} (a) shows the SOA evaluation board with adjustable gain based on the control signal. The gain of SOA can be adjusted by varying the voltage of the control signal. When the voltage of the control signal is varied from 1.5V to 4V, the output power can be adjusted from -60dBm to 5dBm. When the control signal exceeds 4V, the SOA enters the saturation region. Subsequently, we demonstrate the use of the SOA evaluation board to achieve carrier leakage suppression and power difference equalization.

\begin{figure}[!t]
\centering
\includegraphics[width = \linewidth]{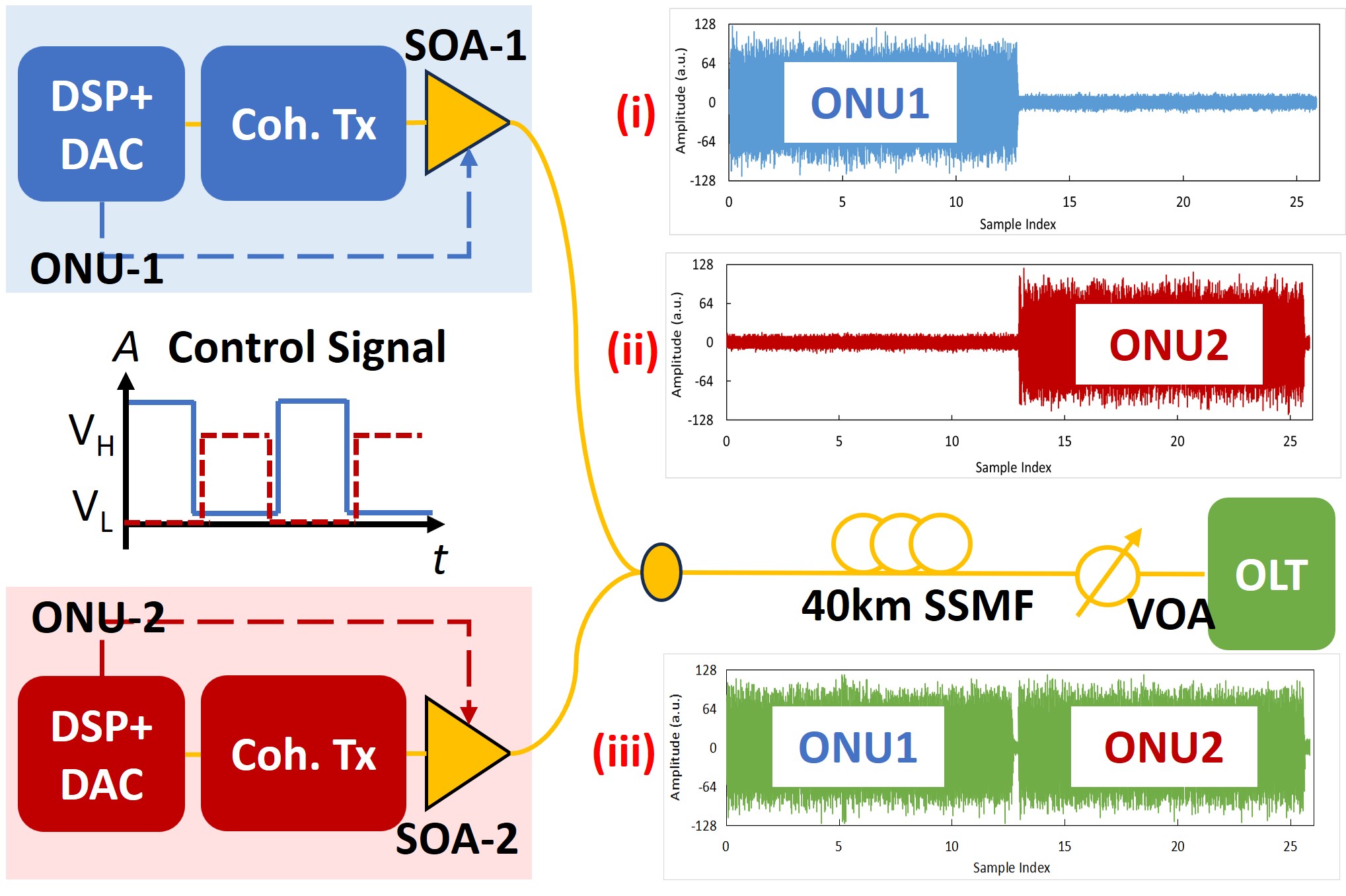}
\caption{The experimental setup of 200Gb/s TDMA based on SOA. Inset (i) and Inset (ii) show the output waveforms of SOA-1 and SOA-2, respectively. Inset (iii) show the TDMA waveform.}
\label{Fig_18}
\end{figure}

\begin{figure}[!t]
\centering
\includegraphics[width =\linewidth]{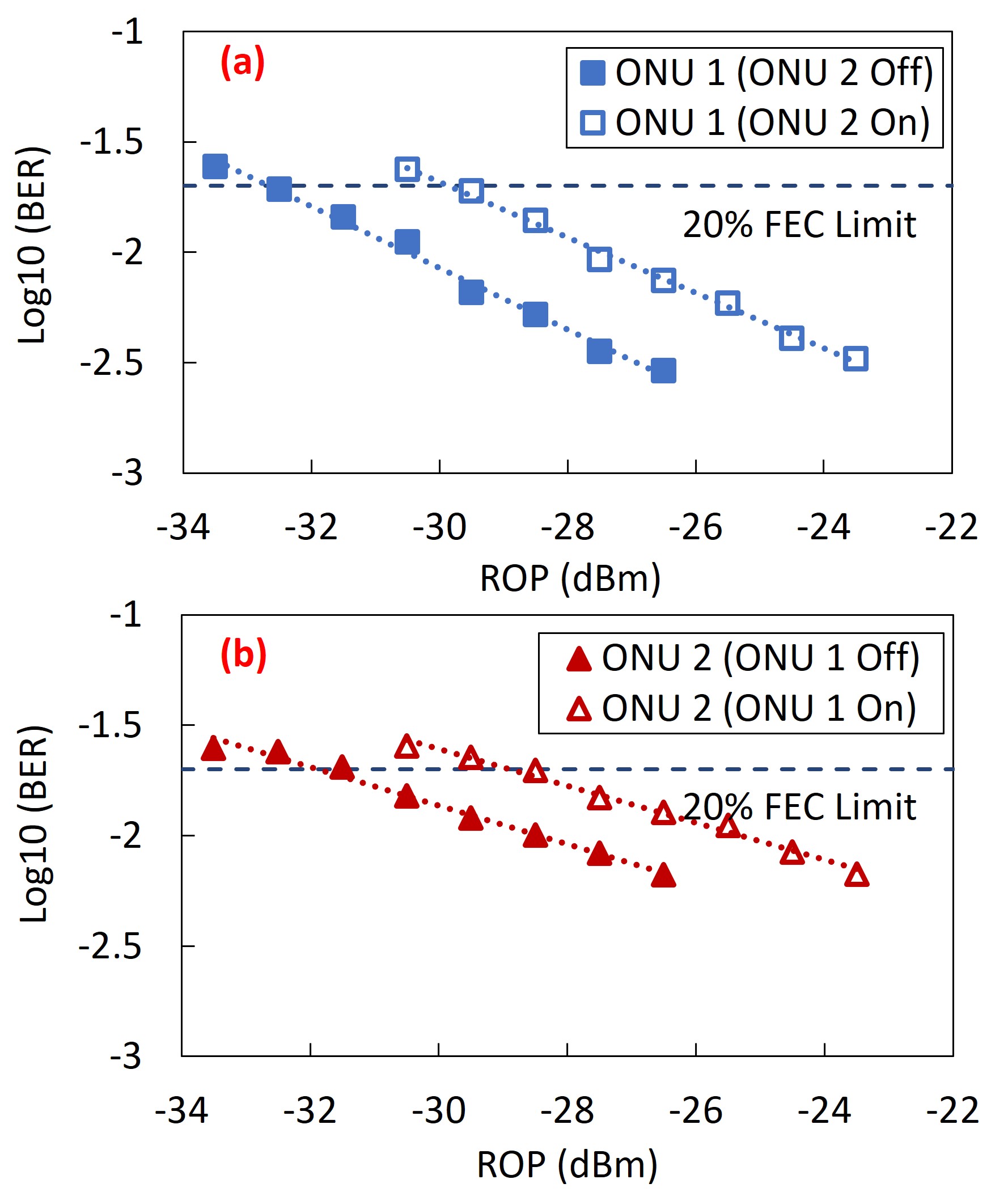}
\caption{BER performance for the ONUs of TDMA based on (a) 40GHz InP and (b) 35GHz LiNbO$_3$ coherent transmitters with SOAs \cite{zhou2026higher}.}
\label{Fig_19}
\end{figure}

\subsection{200Gb/s TDMA based on SOA}
Figure \ref{Fig_18} illustrates the experimental setup of the 200Gb/s SOA-based coherent TDMA. A 40GHz InP coherent transmitter is employed for ONU-1, while a 35GHz LiNbO$_{3}$ coherent transmitter is utilized for ONU-2. With the modulation signal enabled, the output powers of ONU-1 and ONU-2 are $-$13.7dBm and $-$11.5dBm, respectively. When the modulation signal is disabled, the output powers measure $-$44.5dBm and $-$40.5dBm, respectively. As the number of ONUs in the TDMA increases, the performance of individual ONUs will be degraded by carrier leakage from other ONUs. To mitigate this issue, SOAs are implemented to gate the optical paths, thereby enhancing the extinction ratio of the ONUs. As shown in Fig.~\ref{Fig_18}, the SOA control signals for ONU-1 and ONU-2 are inverted. Specifically, SOA-1 is driven by a 348kHz square wave with a high level of 4.75V and a low level of 1.5V, yielding an output power of 2.8dBm after amplification. SOA-2 is driven by a 348kHz square wave with a high level of 4.25V and a low level of 1.5V, also resulting in an output power of 2.8dBm. When the control signal is set to 1.5V, the SOA is switched off, and its output power drops to $-70$dBm (limited by the measurement range of the optical power meter), confirming the high extinction ratio achieved by the SOA gating scheme. Insets (i)-(iii) in Fig. \ref{Fig_18} depict the waveforms of ONU-1, ONU-2, and the composite TDMA signal, respectively. Owing to the effective carrier leakage suppression, the waveforms of ONU-1 and ONU-2 remain mutually non-interfering.

Figure \ref{Fig_19} presents the BER performance of the ONUs utilizing (a) a 40GHz InP coherent transmitter and (b) a 35GHz LiNbO$_3$ coherent transmitter with integrated SOAs. With ONU-2 fully switched off, the required received optical power (ROP) at the 20\% FEC threshold (BER = $2\times 10^{-2}$) for the ONU-1 is approximately $-32.5$dBm, corresponding to an optical power budget of $35.3$dB. Conversely, with ONU-1 fully switched off, the required ROP at the same threshold for the ONU-2 is approximately $-31.5$dBm, yielding a power budget of $34.2$dB. When both ONU-1 and ONU-2 are active, the required ROP at the 20\% FEC limit increases by approximately 3dB. This penalty aligns with theoretical expectations for dual-channel operation, confirming that the ONUs remain mutually non-interfering owing to the high extinction ratio provided by the SOA gating scheme.

\begin{figure}[!t]
\centering
\includegraphics[width = \linewidth]{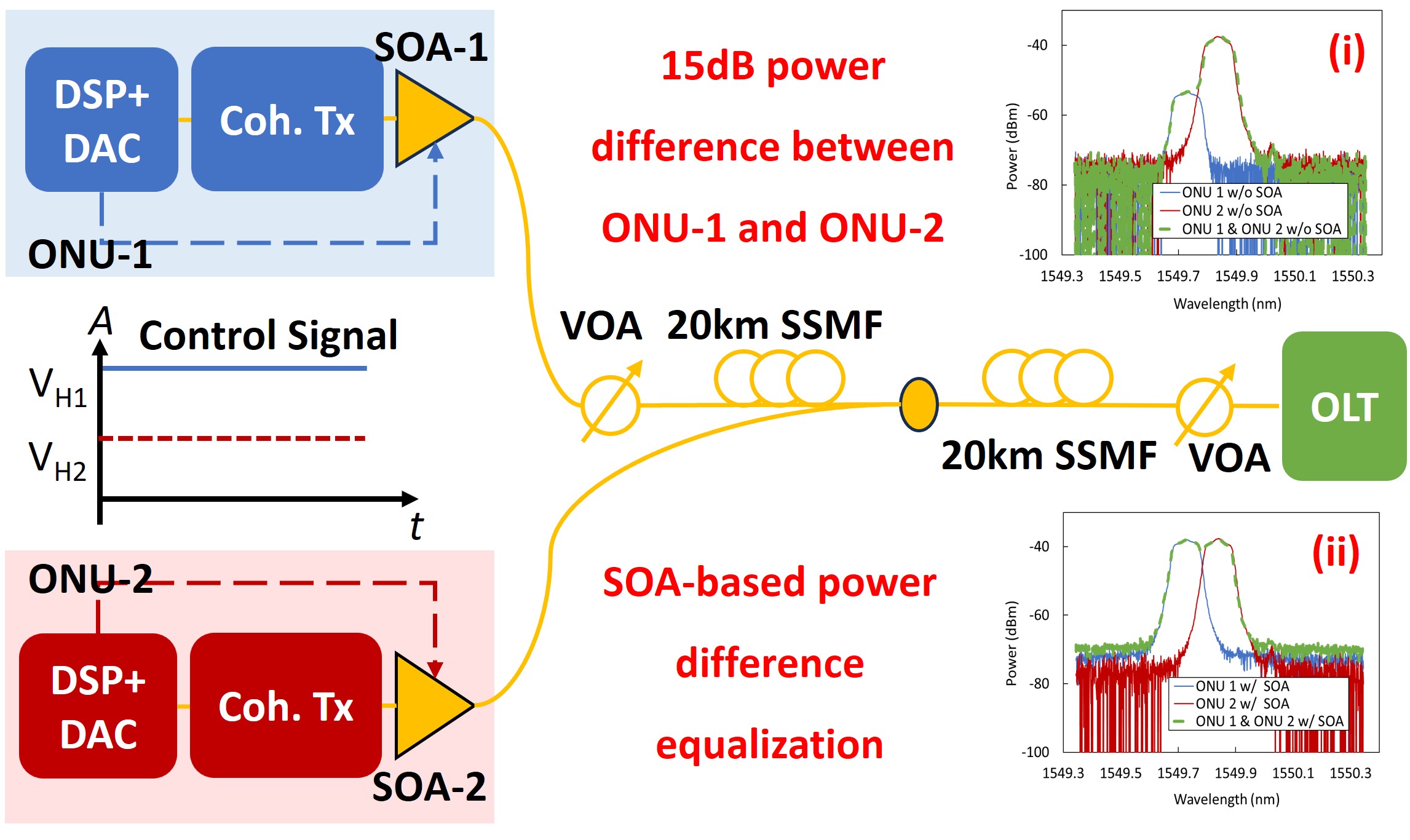}
\caption{The experimental setup of 200Gb/s FDMA based on SOA. Inset (i) shows the spectrum of FDMA signals of two ONUs with 15dB difference of link losses. Inset (ii) shows the spectrum of FDMA signals of two ONUs using power difference equalization based on SOAs}
\label{Fig_20}
\end{figure}

\subsection{200Gb/s FDMA based on SOA}
Figure~\ref{Fig_20} illustrates the experimental setup of the 200Gb/s FDMA based on SOA. The optical signals from ONU-1 and ONU-2 propagate through 40km and 20km of standard single-mode fiber
(SSMF), respectively. To emulate the splitting loss, a variable optical attenuator (VOA) is introduced in the link of ONU-1, intentionally creating a 15dB link loss disparity between the two branches. The SOA control signals serve to dynamically adjust the output power of each ONU. Both SOA-1 and SOA-2 are operated with a constant input power of $-13.5$dBm. When the ONUs emit equal output power, the 15dB link imbalance results in a corresponding 15dB power gap between the two 12.5GBaud subcarriers received at the OLT, as depicted in Inset~(i). Consequently, the low-power subcarrier suffers from degraded SNR and severe crosstalk from its high-power counterpart. To mitigate these impairments, SOA-based power difference equalization is implemented. By applying DC bias voltages of 4.95V and 2.785V to SOA-1 and SOA-2, respectively, the output powers are adjusted to 4.57dBm and $-10.5$dBm. This engineered output difference of approximately 15dB precisely compensates for the link loss asymmetry. As a result, the two 12.5GBaud subcarriers achieve uniform power levels at the OLT, as shown in Inset~(ii). This confirms that the SOA effectively facilitates power difference equalization for ONUs subject to disparate link losses.

\begin{figure}[!t]
\centering
\includegraphics[width = \linewidth]{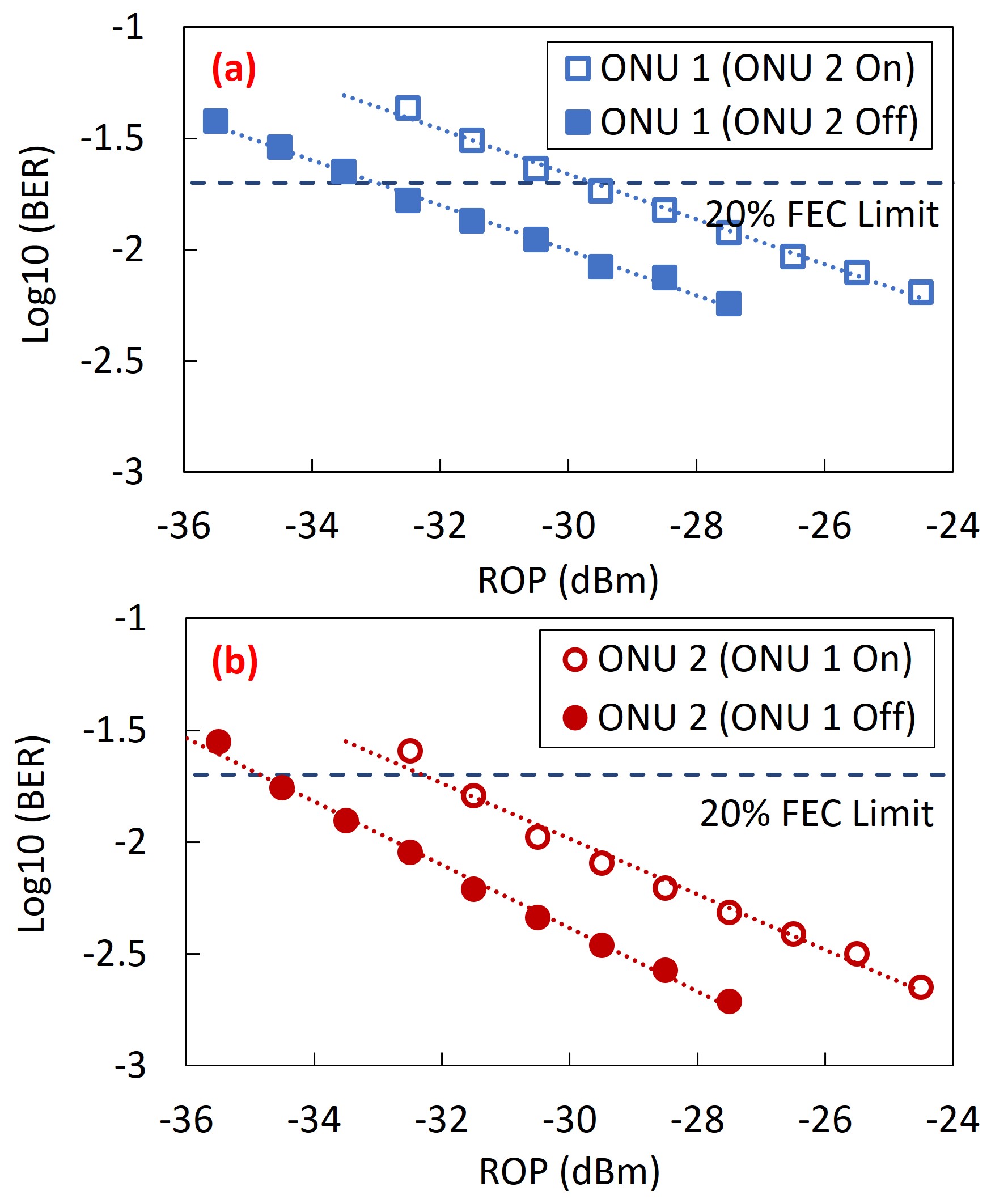}
\caption{BER performance for the ONUs of FDMA based on (a) 40GHz InP and (b) 35GHz LiNbO$_3$ coherent transmitters with SOAs \cite{zhou2026higher}.}
\label{Fig_21}
\end{figure}

Figure~\ref{Fig_21} presents the BER performance of the FDMA employing (a) a 40GHz InP coherent transmitter and (b) a 35GHz LiNbO$_3$ coherent transmitter with integrated SOAs. With ONU-2 fully switched off, the required received optical power (ROP) at the 20\% FEC threshold (BER = $2\times 10^{-2}$) for the ONU-1 is approximately $-33$dBm, yielding an optical power budget of 38dB. This result demonstrates that ONU-1 satisfies the requirements for accommodating large link losses. Conversely, with ONU-1 fully switched off, the required ROP at the same threshold for the ONU-2 is approximately $-34.5$dBm with a 1dB improvement over ONU-1. This enhancement is attributed to ONU-1 suffering from increased chromatic dispersion due to the additional 20km of SSMF and heightened nonlinear distortions resulting from its higher output power relative to ONU-2. When both ONU-1 and ONU-2 are active, the required ROP at the 20\% FEC limit increases by approximately 3dB. This penalty aligns with theoretical expectations for dual-channel operation, confirming minimal mutual interference between the ONUs.

\subsection{Nonlinearity Compensation for SOA}
At high SOA gains, the amplified optical signal suffers from severe nonlinear distortions. To further extend the optical power budget, a nonlinearity compensation algorithm must be employed, with particular consideration given to handling burst-mode signals.

\begin{figure}[!t]
\centering
\includegraphics[width = \linewidth]{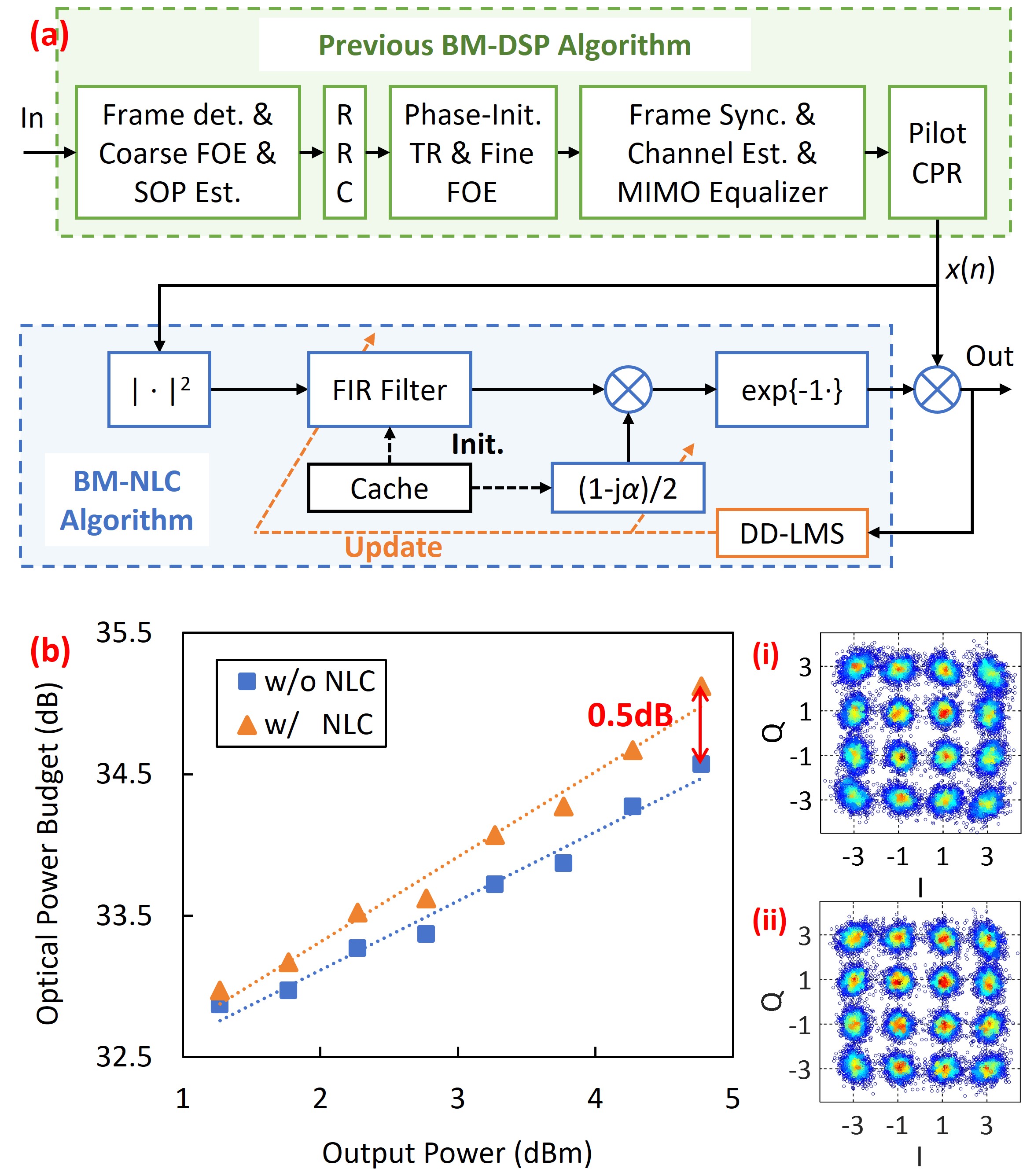}
\caption{BM-DSP algorithm with BM-NLC algorithm for the upstream burst signal processing in coherent PON with SOA. (b) Optical power budget versus output power of SOA for power difference from 11.5dB to 15dB between ONU-1 and ONU-2 with and without BM-NLC. Insets are the recovered 16QAM signal (i) without and (ii) with BM-NLC for ONU-1 at the output power of SOA of 4.77 dBm and ROP of $-$24.5 dBm while ONU-2 is active.}
\label{Fig_7}
\end{figure}

Figure \ref{Fig_7} (a) shows BM-DSP algorithm with the BM-NLC algorithm for the upstream burst signal processing in coherent PON with SOA. The BM-NLC algorithm is designed based on nonlinear model of SOA, which is added after our previous BM-DSP algorithm. The BM-NLC algorithm begins by calculating the instantaneous signal power through a magnitude square operation ($\left|  \cdot \right| ^2$). This power sequence is then processed by an $L$-tap finite impulse response (FIR) filter to approximate the gain exponent. Subsequently, the output of the FIR filter is scaled by the $(1-j\alpha)/2$. This intermediate result is then passed to a complex exponential function ($\text{exp}\left \{- \cdot \right\}$). The resulting complex value is finally multiplied element-wise with the original recovered signal to produce the nonlinearity compensated output. The pre-calculated parameters during the ONU activation are stored in the cache including the tap coefficients of FIR and $\alpha$, which are used for initialization. To realize adaptive parameter tracking after initialization, the tap coefficients of FIR and $\alpha$ can be updated using the decision-directed least mean square error algorithm (DD-LMS) algorithm.

Figure \ref{Fig_7} (b) illustrates the optical power budget performance of the BM-NLC algorithm for SOA-1 across different gains. To simulate varying gain levels, VOA-1 is adjusted to introduce an additional link loss ranging from 11.5 dB to 15 dB, thereby changing the output power of SOA of SOA-1 from approximately 1.27 dBm to 4.77 dBm. The results indicate that higher SOA gain exacerbates nonlinear distortions, making the performance improvement achieved by the proposed BM-NLC algorithm more pronounced under such conditions. For instance, when the output power of SOA is around 1.27 dBm, the nonlinear distortions are relatively minor, resulting in a nearly negligible improvement in the optical power budget. However, at a output power of SOA of approximately 4.77 dBm, the application of the BM-NLC algorithm increases the optical power budget from 34.5 dB to 35 dB, yielding a clear 0.5 dB enhancement. This demonstrates the effectiveness of the proposed BM-NLC algorithm in mitigating SOA-induced nonlinear distortions and expanding the overall optical power budget.

The insets in Fig. \ref{Fig_7} (b) are the recovered 16QAM signal (i) without and (ii) with BM-NLC for ONU-1 at the output power of SOA of 4.77 dBm and ROP of $-$24.5 dBm while ONU-2 is active. Without BM-NLC, the signal suffers from severe SOA-induced nonlinear distortion, causing the constellation points to scatter widely and deviate significantly from the ideal decision points. In contrast, with the proposed BM-NLC, the nonlinear distortion is effectively suppressed, and a markedly larger number of constellation points are tightly clustered around the ideal decision positions, demonstrating a substantial restoration of signal fidelity.

\section{Conclusion}
This work experimentally demonstrates an SOA-based solution for TFDMA coherent PON, effectively tackling two critical challenges: carrier leakage suppression and power difference equalization. By exploiting the fast gating capability of SOAs, we suppress inter-ONU interference in TDMA, while the gain controllability facilitates power balancing for FDMA under disparate link losses. To address the nonlinear distortion induced by high-gain SOA operation, we propose a BM-NLC algorithm integrated into the BM-DSP and yields a 0.5 dB power budget improvement, ultimately achieving a 35 dB optical power budget suitable for B50G-PON deployments.

\vspace{4pt}
\bibliographystyle{unsrt}   
\bibliography{reference}  

\end{document}